\documentclass[twocolumn,aps,prl,showpacs]{revtex4}
\usepackage{graphicx}
\usepackage{amsmath,amssymb}

\newcommand{\ket}[1]{|#1\rangle}

\begin{document}
\title{Minimum Detection Efficiency for a Loophole-Free Atom-Photon
  Bell Experiment}

\author{Ad\'an Cabello} \email{adan@us.es} \affiliation{Departamento
  de F\'{\i}sica Aplicada II, Universidad de Sevilla, E-41012 Sevilla,
  Spain}

\author{Jan-{\AA}ke Larsson} \email{jalar@mai.liu.se}
\affiliation{Matematiska Institutionen, Link\"opings Universitet,
  SE-581 83 Link\"oping, Sweden}


\date{\today}



\begin{abstract}
  In Bell experiments, one problem is to achieve high enough
  photodetection to ensure that there is no possibility of describing
  the results via a local hidden-variable model. Using the
  Clauser-Horne inequality and a two-photon non-maximally entangled
  state, a photodetection efficiency higher than $0.67$ is necessary.
  Here we discuss atom-photon Bell experiments. We show that, assuming
  perfect detection efficiency of the atom, it is possible to perform
  a loophole-free atom-photon Bell experiment whenever the
  photodetection efficiency exceeds $0.50$.
\end{abstract}


\pacs{03.65.Ud,
03.67.Mn,
32.80.Qk,
42.50.Xa}

\maketitle


Forty-three years after Bell's original paper \cite{Bell64}, which
contains what has been described as ``the most profound discovery of
science'' \cite{Stapp75} or, at least, ``one of the greatest
discoveries of modern science'' \cite{Zukowski05}, there is no
experiment testing (the impossibility of) local realism without
invoking supplementary assumptions. All reported Bell experiments,
for instance \cite{FC72,KUW75,Clauser76,FT76,LW76,AGR81,AGR82,
ADR82,OM88,SA88,RT90,OPKP92,KMWZSS95,TBZG98,WJSWZ98,RKVSIMW01,MMBM04},
suffer from so-called ``loopholes.'' It has even been argued that,
``As more time elapses without a loophole-free violation of local
realism, greater should be our confidence on the validity of this
principle [local realism]'' \cite{Santos05}. Beyond this challenge,
quantum information gives us new reasons for performing
loophole-free Bell experiments. There is a link between a true
(``loophole-free'') violation of a Bell inequality and the security
of a family of quantum communication protocols
\cite{Ekert91,SG01,Larsson02}. Specifically, there is an intimate
connection between the existence of a provably secure key
distribution scheme and a true violation of a Bell inequality
\cite{BHK05} (even in the case that quantum mechanics ultimately
fails).

There are two experimental problems that make supplementary
assumptions necessary. The {\em locality loophole} \cite{Bell81}
occurs when the distance between the local measurements is too small
to prevent communication between one observer's measurement choice and
the result of the other observer's measurement. In short, these two
events must be spacelike separated. Massive entangled particles are
extremely difficult to separate \cite{RKVSIMW01}, and high-energy
photons are not appropriate due to the lack of efficient polarization
analyzers. The best candidates for closing the locality loophole are
optical photons, where good polarization analyzers exist and spacelike
separation can be achieved \cite{ADR82,WJSWZ98}. However, thus far,
all Bell experiments with photons suffer from the {\em detection
  loophole} \cite{Pearle70}. The imperfect efficiency of
photodetectors makes the results of all these experiments compatible
with local realistic models. An overall detection efficiency
$\eta>0.67$ is required for two-photon loophole-free Bell
experiments \cite{Eberhard93,LS01}. Single-photon detectors with
more than $0.90$ quantum efficiency already exist, but there are
other difficulties that reduce the overall efficiency to about
$0.30$ or less in practice. Other possible loopholes, e.g., those
related to the subtraction of background counts, will not be
discussed here.

There are some recent proposals as to how to close both the locality
and the detection loopholes simultaneously. One is based on the idea
of achieving entanglement between two separated atoms by preparing
two atom-photon systems and performing a Bell measurement on the
photons which swaps the entanglement to the atoms
\cite{SI03,VWSRVSKW06}. If we accept that the overall measurement
time of the atom is less than $0.5$~$\mu$s, then the two atoms must
be separated at least $150$~m \cite{VWSRVSKW06}. Another proposal is
based on Bell inequalities for two-photon systems prepared in
hyper-entangled states, in which the minimum required photodetection
efficiency is significantly reduced \cite{Cabello06}.

The most promising proposal is the planned Urbana experiment using two
polarization-entangled photons and high-efficiency visible-light
photon counters (VLPCs) \cite{RJAK06,AJRK06}. The actual measured
efficiency of the VLPCs is $0.86$ \cite{AJRK06}. However, after
putting these detectors in a Bell experiment with no less than $60$~m
separation, and considering the background noise, the effective
efficiency could be dangerously close to the minimum required for a
loophole-free experiment ($\eta > 0.75$ with a $0.31\%$ background
noise \cite{Eberhard93}).

Here we show that it is possible to close the detection loophole
with a photodetection efficiency $\eta > 0.50$ by using a single
atom-photon system. Entanglement between the polarization of a
single photon and the internal state of a single trapped atom has
been observed \cite{VWSRVSKW06,BMDM04}. Moreover, a violation of the
Clauser-Horne-Shimony-Holt (CHSH) inequality \cite{CHSH69} with a
cadmium atom and a photon has been reported \cite{MMBM04}. These
experiments, together with new high-efficiency photodetectors,
suggest that a loophole-free atom-photon Bell experiment with a
separation of $150$~m, a perfect detection efficiency for the atom,
and a photodetection efficiency higher than $0.50$ is actually
feasible.

Consider an atom and a photon brought to distant locations. Suppose
$A$ and $a$ are two choices for the observable measured on the atom,
and $B$ and $b$ two choices for the observable measured on the photon.
Each of these observables can only take the values $-1$ or $1$. In
this scenario, a Bell inequality is a necessary constraint imposed by
local realistic theories on the values of a linear combination of
probabilities that can be measured in four different experimental
setups: $(A,B)$, denoting that $A$ is measured on the first particle
and $B$ on the second, $(A,b)$, $(a,B)$, and $(a,b)$.


First, we calculate the minimum detection efficiencies $\eta_A$ and
$\eta_B$ of the atom and the photon detectors, respectively, required
for a loophole-free Bell experiment based on the CHSH inequality
\begin{equation}
|\langle AB \rangle + \langle Ab \rangle + \langle aB \rangle -
\langle ab \rangle| \le 2. \label{CHSH}
\end{equation}
This inequality involves \emph{classical} expectation values of
products of measurement results, and is valid for any local
hidden-variable model with results between $-1$ and $+1$ in the case
of perfect detectors. Quantum expectation values do not obey
(\ref{CHSH}), and the maximum quantum violation is achieved at the
value of $2\sqrt2$ on the left-hand side \cite{Cirelson80}.

In a non-ideal experiment, the expectations are usually calculated by
conditioning on coincidence. In that case, when $\eta_A=\eta_B=\eta$,
it is well known \cite{GM87,Larsson98} that the CHSH inequality
(\ref{CHSH}) must be modified to
\begin{equation}
  |\langle AB \rangle_{\text{coinc}} + \langle Ab
  \rangle_{\text{coinc}}
  + \langle aB \rangle_{\text{coinc}} -
  \langle ab \rangle_{\text{coinc}}| \le \frac4\eta-2. \label{CHSH2}
\end{equation}
This inequality still involves classical conditional expectation
values. Inserting quantum conditional expectation values, the
maximum of the left-hand side is still $2\sqrt2$. Therefore,
(\ref{CHSH2}) is violated only if $\eta>2(\sqrt2-1)\approx0.83$.

Now we want to modify (\ref{CHSH2}) so that it applies to the case
when $\eta_A\neq\eta_B$. To do this we will write, e.g., $A=0$ when
a measurement result is lost due to inefficiency. Thus, assuming
that the rate of non-detections is independent of the measurement
settings, we have
\begin{subequations}
  \begin{align}
    \eta_A = P(A\!\neq\!0)=P(a\!\neq\!0),\\
    \eta_B = P(B\!\neq\!0)=P(b\!\neq\!0).
  \end{align}
\end{subequations}
These are theoretical probabilities which are difficult to extract
from experiment unless \emph{every} experimental run is taken into
account, even those where no detection occurs at either site. Assuming
that detections at one site are independent of detections at the
other, we have
\begin{equation}
  P(A\!=\!B\!=\!0)=(1-\eta_A)(1-\eta_B).
\end{equation}
It is now simple to prove \cite{GM87} that
\begin{equation}
|\langle AB \rangle + \langle Ab \rangle + \langle aB \rangle -
\langle ab \rangle| \le 2-2P(A\!=\!B\!=\!0). \label{CHSH3}
\end{equation}
Furthermore, the probability of a coincidence is $\eta_A\eta_B$ in
this case, and the conditional expectations can be written, e.g.,
\begin{equation}
  \langle{}AB\rangle_{\text{coinc}}=\frac1{\eta_A\eta_B}\langle AB
  \rangle.
\end{equation}
Thus, (\ref{CHSH3}) can be written
\begin{equation}
  \begin{split}
    |\langle AB \rangle_{\text{coinc}} &+ \langle Ab
    \rangle_{\text{coinc}} + \langle aB \rangle_{\text{coinc}} -
    \langle ab \rangle_{\text{coinc}}| \\&
    \le \frac{2\eta_A+2\eta_B-2\eta_A\eta_B}{\eta_A\eta_B}
    = \frac2{\eta_A}+\frac2{\eta_B}-2. \label{CHSH4}
  \end{split}
\end{equation}
This inequality is a generalization of (\ref{CHSH2}). Inserting the
maximum quantum value $2\sqrt2$ in the left-hand side yields a bound
for $\eta_B$ as a function of $\eta_A$. In brief, inequality
(\ref{CHSH4}) is violated only if
\begin{equation}
  \label{eq:2}
  \eta_B>\frac{\eta_A}{(\sqrt2+1)\eta_A-1}.
\end{equation}
In the special case when $\eta_A=1$, there is a violation only if
$\eta_B>1/\sqrt2\approx0.71$.


Although the CHSH inequality and the Clauser-Horne (CH) inequality are
equivalent in the ideal case \cite{CH74}, the CH inequality is
violated by quantum mechanics as soon as
$\eta=\eta_A=\eta_B>\frac23\approx0.67$ \cite{Eberhard93,LS01}. That
is, even when $0.67<\eta<0.83$. We therefore expect to be able to
lower the above $0.71$ bound using the CH inequality in an atom-photon
experiment. The CH inequality can be written
 \begin{equation}
  \begin{split}
    P(A\!=\!B\!=\!1)+P(A\!=\!b\!=\!1)+P(a\!=\!B\!=\!1)& \\
    -P(a\!=\!b\!=\!1)-P(A\!=\!1)-P(B\!=\!1)& \le 0,
  \end{split}
  \label{CH}
\end{equation}
where $P(A\!=\!1)$ is the probability that $A=1$ without a
corresponding detection being required at the other site. This
inequality has the same status as (\ref{CHSH}), and relates classical
probabilities from a local hidden-variable model. The quantum
probabilities do not obey the CH inequality and the maximum quantum
value of the left-hand side is $\sqrt2-1$. However, as we shall see,
the probabilities in the CH inequality scale differently with the
efficiency, so the maximum quantum value does not coincide with the
minimum efficiency for which there is violation of the inequality.

Indeed, if
\begin{equation}
3\eta_A \eta_B-\eta_A-\eta_B >0, \label{veinticinco}
\end{equation}
there are quantum states and local observables violating (\ref{CH}).
For instance, we can use the state \cite{LS01}
\begin{equation}
  \begin{split}
    \ket{\psi}=&C\left\{\left[1-2\cos(\theta)\right])\ket{0_{a}0_{b}}\right.
    \\&\left.+\sin(\theta)\left(\ket{0_{a}1_{b}}
      +\ket{1_{a}0_{b}}\right)\right\}, \label{veintiseis}
  \end{split}
\end{equation}
and the local observables $A$ and $B$, defined from the local
observables $a$ and $b$, respectively, by
\begin{subequations}
  \begin{align}
    \begin{pmatrix}
      \ket{0_{A}}\\\ket{1_{A}}
    \end{pmatrix}
    &=
    \begin{bmatrix}
      \cos(\theta)&-\sin(\theta)\\\sin(\theta)&\cos(\theta)
    \end{bmatrix}
    \begin{pmatrix}
      \ket{0_{a}}\\\ket{1_{a}}
    \end{pmatrix},
    \label{veintisiete} \\
    \begin{pmatrix}
      \ket{0_{B}}\\\ket{1_{B}}
    \end{pmatrix}
    &=
    \begin{bmatrix}
      \cos(\theta)&-\sin(\theta)\\\sin(\theta)&\cos(\theta)
    \end{bmatrix}
    \begin{pmatrix}
      \ket{0_{b}}\\\ket{1_{b}}
    \end{pmatrix}.
    \label{veintiocho}
  \end{align}
\end{subequations}
Taking the efficiencies into account, and using
$\epsilon=\tan\frac\theta2$ and $K=\sin^2\theta$, we arrive at the
\emph{quantum} probabilities
\begin{subequations}
  \begin{align}
    P(A\!=\!B\!=\!1)&=K\eta_A\eta_B > 0,\label{veintinueve}\\
    P(A\!=\!b\!=\!1)&=K\eta_A\eta_B,\label{treintayuno}\\
    P(a\!=\!B\!=\!1)&=K\eta_A\eta_B,\label{treintaydos}\\
    P(a\!=\!b\!=\!1)&=0,\label{treinta}\\
    P(A\!=\!1)&=K\eta_A(1+\epsilon^2),\label{treintaycinco}\\
    P(B\!=\!1)&=K\eta_B(1+\epsilon^2). \label{treintayseis}
  \end{align}
\end{subequations}
For this state and these observables, the left-hand side of the CH
inequality (\ref{CH}) is
\begin{equation}
K\left[3\eta_A \eta_B-\eta_A-\eta_B-
\epsilon^2\left(\eta_A+\eta_B\right)\right]. \label{treintaysiete}
\end{equation}
The interesting point is that, when $3\eta_A \eta_B-\eta_A-\eta_B>0$,
we simply choose $\epsilon>0$ such that $3\eta_A
\eta_B-\eta_A-\eta_B>\epsilon^2\left(\eta_A+\eta_B\right)$ (we need to
choose $\epsilon\neq0$ otherwise $K=0$). Using this value of
$\epsilon$, or rather $\theta=2\arctan\epsilon$, we can construct a
quantum state $\ket{\psi}$ and four local observables $A$, $a$, $B$,
and $b$ such that the CH inequality (\ref{CH}) is violated.

We continue by proving that a violation can be obtained \emph{only
if} (\ref{veinticinco}) is satisfied. We again use the notation
$A=0$ to denote when a measurement result is lost due to
inefficiency.  We condition on detection on one side, and write the
conditional probabilities as, e.g.,
\begin{equation}
  P_{\text{detect}}(A\!=\!1)=\frac1{\eta_A}P(A\!=\!1).
\end{equation}
We also, again, assume that detections at one site are independent
of detections at the other. Then, the probabilities must satisfy the
following trivial inequalities:
\begin{equation}
  \begin{split}
    P(A\!=\!B\!=\!1)
    \le \eta_A\eta_B\displaystyle\min_{X=A,B}P_{\text{detect}}(X\!=\!1),
  \end{split}   \label{catorce}
\end{equation}
\begin{equation}
  \begin{split}
    P(A\!=\!b\!=\!1)&-P(A\!=\!1)
    \\&\le(\eta_B-1)\eta_A\displaystyle\min_{X=A,B}P_{\text{detect}}(X\!=\!1),
  \end{split}
\end{equation}
\begin{equation}
  \begin{split}
    P(a\!=\!B\!=\!1)&-P(B\!=\!1)\\
    &\le(\eta_A-1)\eta_B\displaystyle\min_{X=A,B}P_{\text{detect}}(X\!=\!1),
  \end{split}
\end{equation}
and
\begin{equation}
  -P(a\!=\!b\!=\!1) \le 0. \label{quince}
\end{equation}
Therefore, \emph{only} assuming that detections at one site are
independent of detections at the other, the left-hand side of the CH
inequality (\ref{CH}) must obey
\begin{equation}
  \begin{split}
    P(A\!=&\!B\!=\!1)+P(A\!=\!b\!=\!1)+P(a\!=\!B\!=\!1) \\
    -P(&a\!=\!b\!=\!1)-P(A\!=\!1)-P(B\!=\!1)\\
    &\le \left(3\eta_A \eta_B-\eta_A-\eta_B\right)\min_{X=A,B}
    P_{\text{detect}}(X\!=\!1).
  \end{split}
\label{veintidos}
\end{equation}
The left-hand side of (\ref{veintidos}) can be positive only if
$3\eta_A \eta_B-\eta_A-\eta_B>0$, so the CH inequality (\ref{CH})
can only be violated if this is the case.

Summing up, the CH inequality (\ref{CH}) is violated if and only if
inequality (\ref{veinticinco}) is satisfied, or
\begin{equation}
\eta_B > \frac{\eta_A}{3\eta_A-1}.
\end{equation}
Note especially that the CH inequality (\ref{CH}) is violated if and
only if
\begin{align}
  \eta_B > \frac{1}{2},\;&\text{ when }\eta_A=1, \\
  \eta_B > \frac{2}{3},\;&\text{ when }\eta_B=\eta_A.
\end{align}
Therefore, closing the detection loophole with an atom-photon system
using the CH inequality, and assuming perfect detection efficiency
of the atom, requires a minimum photodetection efficiency of $0.50$
(vs $\eta_B = 0.67$ for the photon-photon case
\cite{Eberhard93,LS01}).

Choosing $\ket{\psi}$ to be a maximally entangled state and assuming
that $\eta_A=1$, the minimum $\eta_B$ required for a loophole-free
Bell experiment coincides with that previously calculated for the CHSH
inequality.

So far, we have assumed that only the pairs prepared in the entangled
state contribute to the counting rates, and that local measurements
are perfect. In order to take into account deviations from that ideal
case, we now introduce {\em background noise} (see Fig.~1.),
\begin{figure}[b]
\center
\includegraphics[width=\linewidth]{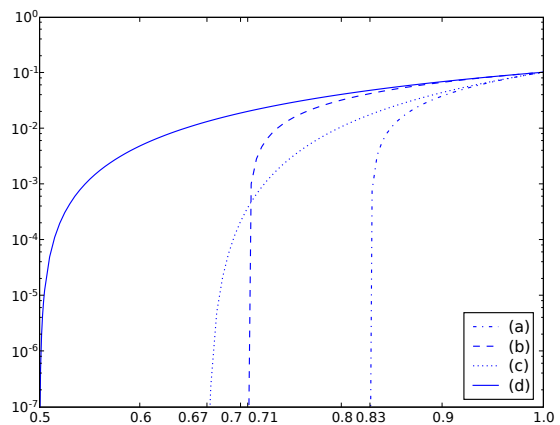}
\caption{Maximum affordable background noise for a loophole-free Bell
  experiment as a function of the photodetection efficiency $\eta_B$.
  (a) using the CHSH inequality with $\eta_A=\eta_B$, (b) using the
  CHSH inequality with $\eta_A=1$, (c) using the CH inequality with
$\eta_A=\eta_B$, and (d) using the CH inequality with $\eta_A=1$. The
cases (a) and (c) are appropriate in a photon-photon experiment and
the cases (b) and (d) are appropriate in an atom-photon experiment.}
\label{figure}
\end{figure}
as in \cite{Eberhard93}, and we have numerically found the maximum
affordable background noise for a loophole-free Bell experiment as a
function of the photodetection efficiency, in four relevant cases,
combining the usage of the CH or the CHSH inequalities with a
photon-photon or an atom-photon experiment.  In the case of two
photons, our calculations agree with those in \cite{Eberhard93}. The
detailed calculations will be presented elsewhere.

The main result can be summarized as follows: Using an atom-photon
system, and assuming perfect detection efficiency of the atom (as is
usually the case in actual experiments), the minimum photodetection
efficiency required for a loophole-free Bell experiment can be
lowered to $0.50$ in the absence of noise (vs $\eta > 0.67$ for the
photon-photon case), and to $0.58$ for a $0.31\%$ background noise
(vs $\eta > 0.75$ for the photon-photon case). This result suggests
a new approach for performing a loophole-free Bell experiment.


\begin{acknowledgments}
  The authors thank J.B. Altepeter, P. Grangier, P. G. Kwiat, R.
  Rangarajan, H. Weinfurter, and M. \.{Z}ukowski for useful
  conversations. A.C. thanks M. Bourennane for his hospitality in
  Stockholm University, and acknowledges support from the Spanish MEC
  Project No. FIS2005-07689, and the Junta de Andaluc\'{\i}a
  Excellence Project No. FQM-2243. J.-\AA.L. acknowledges support from
  the Swedish Science Council.
\end{acknowledgments}


{\em Note added in proof.}---After submitting this manuscript, we
have become aware that some results presented here have been
independently derived by Brunner {\em et al.} \cite{BGSS07}.

\end{document}